# GPU in the Blind Spot: Overlooked Security Risks in Transportation

**Sefatun-Noor Puspa**
Ph.D. Student, Glenn Department of Civil Engineering
Clemson University, Clemson, South Carolina, 29634
Email: spuspa@g.clemson.edu

**Mashrur Chowdhury, Ph.D., P.E.**
Eugene Douglas Mays Chair of Transportation, Glenn Department of Civil Engineering
Clemson University, Clemson, South Carolina, 29634
Email: mac@clemson.edu

Word Count: 5873 words + 3 table (250 words per table) = 6623 words

*Submitted [08/01/2025]*

**ABSTRACT**
Graphics processing units (GPUs) are becoming an essential part of the intelligent transportation system (ITS) for enabling video-based and artificial intelligence (AI) based applications. GPUs provide high-throughput and energy-efficient computing for tasks like sensor fusion and roadside video analytics. However, these GPUs are one of the most unmonitored components in terms of security. This makes them vulnerable to cyber and hardware attacks, including unauthorized crypto mining. This paper highlights GPU security as a critical blind spot in transportation cybersecurity. To support this concern, it also presents a case study showing the impact of stealthy unauthorized crypto miners on critical AI workloads, along with a detection strategy. We used a YOLOv8-based video processing pipeline running on an RTX 2060 GPU for the case study. A multi-streaming application was executed while a T-Rex crypto miner ran in the background. We monitored how the miner degraded GPU performance by reducing the frame rate and increasing power consumption, which could be a serious concern for GPUs operating in autonomous vehicles or battery-powered edge devices. We observed measurable impacts using GPU telemetry (nvidia-smi) and nsight compute profiling, where frame rate dropped by 50 percent, and power usage increased by up to 90%. To detect, we trained lightweight classifiers using extracted telemetry features. All models achieved high accuracy, precision, recall, and F1-score. This paper raises urgent awareness about GPU observability gaps in ITS and offers a replicable framework for detecting GPU misuse through on-device telemetry.

*Puspa et al.*

**INTRODUCTION**

Graphics Processing Units (GPUs) have become essential components in modern transportation systems in enabling intelligent infrastructure and vehicles. From smart roadside units (RSUs) equipped with video analytics to traffic surveillance cameras and advanced driver-assistance systems (ADAS), GPUs empower real-time processing of high-resolution data streams. Their parallel computing capabilities support critical tasks such as object detection, license plate recognition, and traffic flow analysis. The integration of GPUs in both stationary and mobile platforms has enhanced the responsiveness and intelligence of transportation networks, contributing to smarter and safer roadways *(1)(2)*.

This evolution represents a significant shift from traditional CPU-bound systems to GPU-accelerated edge intelligence. Modern platforms such as NVIDIA Jetson AGX Orin, Jetson Xavier, and DRIVE PX have been widely adopted in autonomous vehicle prototypes and edge devices *(3)*, offering substantial improvements in performance per watt for machine learning workloads. Even discrete GPUs like NVIDIA's RTX series are increasingly used in research-grade automotive testbeds for rapid prototyping and on-board AI inference. These platforms enable complex computations at the edge, reducing latency, lowering bandwidth consumption, and increasing overall system efficiency *(3)(4)*.

Despite the widespread deployment of GPUs in transportation infrastructure, security models have not kept pace with this technological shift. Traditional cybersecurity frameworks predominantly focus on protecting the operating system, CPU-bound processes, and network-level communication. As a result, GPU workloads, especially those executed at the kernel level or through vendor-specific APIs like CUDA or OpenCL, often fall outside the visibility of conventional endpoint detection and response (EDR) tools *(5)(6)*. This blind spot creates a critical vulnerability in modern intelligent transportation systems.

GPU processes can be exploited without triggering standard alerts, which allows attackers to stealthily run malicious workloads such as unauthorized cryptocurrency miners. For example, a stealth crypto miner embedded in a roadside AI unit or autonomous vehicle can quietly drain computational resources, leading to significant degradation in system performance. This includes reduced frames per second (FPS) in vision pipelines, thermal throttling, increased power drawing, and even delayed safety-critical responses. These safety-critical applications include pedestrian detection *(7)* or collision avoidance *(8)*, which can be affected by the stealthy miner without being flagged by traditional security mechanisms. Such threats pose direct risks to both passenger safety and infrastructure integrity.

This paper addresses the missing link between GPU deployment and GPU security in transportation systems. First, it highlights emerging threats that specifically exploit GPU resources in ITS. Next, it reviews current detection methods, including profiling tools and telemetry analysis, that have potential for GPU-focused anomaly detection. We present a case study in the paper, demonstrating the feasibility of detecting unauthorized GPU miners using onboard telemetry and performance counters. Finally, the paper gives directions for future research, secure deployment strategies, and the need for standardized GPU-aware security models across transportation platforms.

**GPU IN MODERN TRANSPORTATION SYSTEMS**

In modern intelligent transportation systems (ITS), GPU plays a vital role in powering a wide range of critical applications that require real-time, high-throughput computation. At the core of many ITS deployments is computer vision, where GPU accelerates object detection, traffic violation monitoring, and vehicle tracking using deep learning models. Whether deployed in roadside cameras or mounted on infrastructure at intersections, GPUs process high-resolution camera feeds to identify pedestrians, recognize license plates, classify vehicle types, and detect red-light running or speeding incidents *(9)(10)*. These tasks demand rapid inference speeds and parallel computation to ensure minimal latency capabilities that traditional CPU-bound systems cannot provide *(11)*. Moreover, advanced video analytics, including motion trajectory analysis and multi-object tracking, benefit from the parallelism of GPU architecture, which makes them indispensable for cities. These characteristics of the GPU help to automate traffic management or improve situational awareness on busy roadways.

GPUs also empower ITS in real-time sensor fusion for cooperative perception at intersections and other high-traffic zones. In these environments, multiple data sources such as LiDAR, radar, video feeds

*(12)*, and V2X messages *(13)* must be fused to form a comprehensive, low-latency environmental model. This cooperative perception is vital for early hazard detection and coordinated decision-making between smart infrastructure and connected vehicles. GPUs facilitate this fusion by running parallel algorithms and real-time AI-based perception models. Their ability to handle heterogeneous sensor data at high throughput enables smarter, safer interactions at intersections, especially where human visibility or vehicle line-of-sight is limited *(14)*.

Within vehicles, especially those supporting Advanced Driver Assistance Systems (ADAS) and autonomous driving features, GPUs are central to decision-making and perception. Tasks such as simultaneous localization and mapping (SLAM) *(15)*, 3D semantic segmentation, lane detection, and trajectory planning require not just rapid computation but also low-power acceleration, an area where modern automotive-grade GPUs excel *(16)*. Neural networks used for obstacle avoidance, path prediction, and driver monitoring rely heavily on GPU parallelism for real-time inference. High-end autonomous platforms use convolutional neural networks (CNNs), recurrent neural networks (RNNs), and transformers for continuous sensing and control *(17)*. These AI workloads are too intensive for CPUs or microcontrollers alone, making onboard GPUs essential for both safety and performance *(18)*.

To meet the diverse computational needs of these ITS applications, a variety of GPU-based hardware platforms are employed across the transportation ecosystem. The NVIDIA Jetson Orin platform has become standard in edge-deployed smart roadside units (RSUs) and traffic cameras, providing a compact, low-power solution with robust AI acceleration. Jetson Orin can run full perception pipelines locally, enabling infrastructure to respond to dynamic traffic conditions in real time. For autonomous vehicles, especially shuttles or industrial fleets, NVIDIA's AGX platform offers an automotive-grade solution designed for mission-critical AI tasks *(19)*, featuring redundant safety mechanisms and deep learning optimization*(18)*. Additionally, RTX-series GPUs commonly used in consumer desktops are increasingly adopted in development kits and traffic analytics systems, where researchers and engineers need flexible, high-performance computing for prototyping or running pilot deployments. These GPUs offer powerful compute and Tensor Core acceleration, making them suitable for testing perception and planning algorithms before transitioning to embedded or production-grade platforms.

Despite their capabilities, GPUs deployed in field-based ITS systems face a unique set of constraints. These edge devices must operate autonomously and reliably in outdoor or vehicular environments with limited connectivity, power, and maintenance support. For example, an RSU might be mounted on a traffic pole without constant network access or physical security, making remote diagnostics or software updates difficult. This necessitates that the GPU-based system not only performs its primary tasks but also be resilient to faults and adaptive to environmental changes. Furthermore, these edge-deployed GPUs are often underutilized relative to their computer capacity, running on intermittent workloads depending on traffic density and time of day. This underutilization, combined with the lack of real-time monitoring and fine-grained resource auditing, creates an attractive attack surface for malicious or unauthorized usage. For instance, a stealthy crypto miner could exploit idle GPU cycles *(20)* without triggering conventional alarms, leading to overheating, degraded AI performance, and compromised safety, a risk largely overlooked in traditional ITS security models *(21)*. Thus, while GPUs significantly elevate the capabilities of ITS systems, their powerful, low-visibility nature also introduces new security challenges that demand closer attention from both researchers and practitioners.

**GPU SECURITY RISKS**

The misuse of GPU resources in intelligent transportation systems (ITS) is no longer a hypothetical concern; it is an emerging security challenge with safety and operational consequences. One of the most immediate risks is unauthorized crypto mining. In GPU-accelerated platforms such as smart roadside units (RSUs), traffic analytics boxes, or in-vehicle systems, idle GPU cycles can be hijacked to run crypto mining algorithms without the operator's knowledge *(20)*. This form of "cryptojacking" drains power, causes excess heat generation, and degrades performance, particularly during off-peak hours. Such attacks have been demonstrated in academic studies, where GPU-based miners can be embedded within machine learning pipelines and escape traditional endpoint detection systems. Other studies further show how GPU-based crypto miners can be injected into poorly secured edge devices, converting computational infrastructure into silent, energy-intensive mining nodes.

Beyond mining, GPUs open up a new vector for covert computation. Unlike CPUs, which are typically subject to process-level monitoring and host-based intrusion detection, GPU workloads often execute in isolated memory spaces using separate kernel streams *(22)(23)*. This creates a blind spot for most IT security tools. Attackers can exploit this by embedding malicious GPU kernels into otherwise benign applications, for instance, hijacking the Tensor cores to conduct password cracking, data exfiltration, or AI model inversion, all without involving the core processor. This form of stealth computation is particularly dangerous in ML environments because of the lack of mature runtime security for inference engines and accelerators. The covert nature of these tasks, coupled with the high throughput of modern GPUs, makes them ideal platforms for stealthy attacks that escape detection entirely *(24)*.

Another underexplored but dangerous risk lies in denial-of-service (DoS) *(25)* scenarios caused by GPU kernel abuse. A malicious actor can write kernels that consume all shared memory, loop infinitely, or saturate thread blocks, leading to congestion in the GPU's execution pipeline *(26)*. On real-time transportation systems, this is catastrophic. For example, a smart camera processing intersection video feeds may begin to drop frames or output delayed detections if its GPU is locked in an unauthorized loop. Autonomous driving stacks could experience blocked inference pipelines *(27)*, causing perceptual or decision lag. These kernel-level DoS conditions can disrupt critical scheduling, cause device reboots, or stall safety systems in edge environments.

Closely tied to these issues is the phenomenon of thermal degradation. Unlike cloud GPUs housed in climate-controlled data centers, GPUs in roadside or in-vehicle deployments are exposed to environmental temperature variations and often rely on passive cooling *(28)*. Continuous high-load operations, especially unauthorized ones like crypto mining, cause sustained thermal stress that can degrade GPU silicon over time, reduce reliability, and even trigger thermal throttling that slows down safety-critical workloads *(29)*. While thermal throttling protects hardware, it leads to degraded AI performance, potentially impacting time-sensitive perception and planning tasks. In systems meant to operate for years without maintenance, this degradation presents a real-world reliability concern, especially in under-maintained roadside infrastructure *(29)*.

These vulnerabilities translate into concrete failures in ITS operations. A GPU compromised by rogue code or mining activity may begin to lag in perception tasks such as object detection from video or radar input. As frame rates drop or detection latency increases, important visual cues like a pedestrian entering a crosswalk or a vehicle running a red light can be missed *(30)*. Similarly, in cooperative perception systems that fuse camera and LiDAR data, GPU lag can lead to misaligned fusion, producing a false or stale view of the environment. In connected vehicle contexts, GPU-bound delays in V2X message fusion or validation could disrupt real-time safety coordination *(31)*, leading to missed alerts or incorrect actions at intersections.

A key part of this problem is that most edge systems lack security tools with GPU-level visibility. Traditional security agents focus on file systems, network behavior, and CPU process trees. However, GPU workloads operate through separate APIs (e.g., CUDA), run in separate memory, and often don't generate audit trails accessible to host-based monitoring tools. Tools like NVIDIA Management Library (NVML) or CUPTI (for GPU telemetry) are required to observe power usage, memory allocation, and kernel execution, but these tools are often disabled or not installed in production environments due to IT policies or driver limitations. Research-grade installations may leave CUPTI disabled to reduce overhead, and industrial deployments may not expose NVML due to a lack of software integration. This creates a critical blind spot, where GPU abuse can persist indefinitely without visibility or alerts.

The entry points for attackers are diverse and feasible for GPUs. In many deployments, systems rely on remote management or over-the-air (OTA) software updates, both of which can be exploited. If an attacker phishes a system administrator or compromises the OTA pipeline, they can insert malicious GPU binaries during what appears to be a routine update. Worse yet, many roadside units and cameras are physically accessible or insecurely mounted. A USB drive dropped into a roadside cabinet or an exposed debug port on a pole-mounted RSU can provide physical access to the system's storage or bootloader, enabling payload injection. Once deployed, the GPU kernel code can remain dormant or lightly loaded, avoiding detection while continuously mining or degrading system function *(32)*.

In conclusion, the misuse of GPU resources in transportation infrastructure represents a form of silent safety degradation. Unlike network attacks that generate alerts or malware that crashes systems, GPU abuse hides in plain sight, gradually slowing perception, introducing errors, and accelerating hardware aging. These effects, though subtle, compromise the responsiveness and trustworthiness of ITS systems. As the field moves toward increasingly GPU-reliant architectures, whether for AVs, RSUs, or infrastructure analytics, it is essential that researchers and operators treat GPU runtime integrity as a first-class security concern. This means investing in GPU-aware monitoring tools, hardening update channels, and establishing behavioral baselines for edge AI workloads to ensure both safety and resilience.

**REVIEW OF EXISTING DETECTION METHODS FOR GPU INCURSION**
While GPUs have become indispensable for real-time AI in transportation systems, detection mechanisms to secure them remain underdeveloped. Unlike CPUs, which benefit from decades of robust monitoring infrastructure, GPUs lack standardized, field-viable security research. This section surveys the state of current detection techniques and their limitations in real-world intelligent transportation settings.

**Existing Detection Techniques**
Several approaches have been explored for detecting malicious or unauthorized GPU activity, often repurposed from CPU-centric or data center contexts:

CPU Hardware Performance Counters (HPCs) are frequently used to detect anomalies in instruction flow, cache usage, and branch prediction features commonly exploited by malware. However, these counters are limited to CPU execution and offer no visibility into GPU-side operations such as CUDA kernel launches, memory transfers, or shared memory utilization *(33)*. Thus, they are ineffective for detecting GPU-resident threats like crypto miners or covert AI workloads running in edge devices.

External Sensors, including power side-channel analysis tools and electromagnetic (EM) probes, have been used to detect unauthorized GPU behavior in lab environments. For instance, Xiao et al. *(34)* demonstrated power trace-based detection of hidden workloads by analyzing EM leakage patterns. While effective in controlled setups, these methods are impractical for deployment in fielded systems like roadside units (RSUs) or in-vehicle compute boxes due to cost, complexity, and environmental noise sensitivity.

Power-Based Monitoring via APIs such as NVIDIA's NVML (NVIDIA Management Library) or tools from the RAPIDS suite provides access to metrics like GPU power draw, memory usage, and thermal readings. While theoretically useful for detecting sustained abnormal loads from stealthy mining or inference hijacking, such telemetry access is frequently restricted on enterprise-managed or production edge systems. Additionally, API-based monitoring is often blocked for non-root processes or sandboxed environments due to security policies, further limiting its real-world utility.

**Absence of Continuous Runtime Solutions**
Current transportation deployments lack continuous, runtime GPU data access. Security policies within ITS architecture and vehicle control systems often omit the GPU entirely, treating it as an opaque co-processor rather than an actively monitored subsystem *(35)*. This oversight leaves systems vulnerable to persistent stealth threats that degrade performance or manipulate sensor inference without raising alerts *(36)*.

**CASE STUDY: ON-DEVICE DETECTION OF GPU CRYPTOMINING**
Graphics processing units (GPUs) are becoming a critical component in the operation of intelligent transportation systems. They are used for real-time video analytics, object detection, and sensor data processing that enable safe and responsive AI behavior on the road. However, despite their growing role, GPU activity often remains unmonitored, leaving them exposed to silent misuse. One such threat is unauthorized crypto mining, where malicious software hijacks the GPU's computing resources for profit without disrupting visible operations *(20)*. These miners can operate in the background of safety-critical applications, draining power and slowing down performance in ways that may go unnoticed.

In this case study, we examine how the presence of a crypto miner affects a GPU-based perception workload and whether this misuse can be identified using only on-device monitoring tools. We set up a video streaming application using YOLOv8 to simulate a roadside AI system, and introduced a T-Rex crypto miner configured to mine Ravencoin using the KawPoW algorithm. Both tasks ran simultaneously on the same GPU, mimicking a stealth attack that could occur in a real-world edge deployment. The goal was to observe changes in performance and GPU behavior under malicious load and to test if these changes could be used to detect the threat reliably without external sensors or privileged system access.

**Experimental Setup**
To simulate a transportation-focused edge computing environment, we designed a desktop testbed that mimics the conditions of a roadside AI unit. The experimental setup used for emulating a roadside GPU environment is summarized in **Table 1**.

**Table 1: Experimental Setup Summary**

| Component | Description |
| --- | --- |
| CPU | AMD Ryzen 5 3600 |
| GPU | NVIDIA RTX 2060 Super |
| RAM | 8 GB |
| OS | Windows 10 Pro |
| Benign Workload | Multi-stream YOLOv8 (video detection) |
| Malicious Workload | T-Rex Miner (KawPoW for Ravencoin) |
| Telemetry Tools | nvidia-smi, Nsight Compute |

For the benign workload, we used a multi-streaming video application running YOLOv8 to perform object detection on 1080p video feeds. This mirrors the perception workload common in roadside cameras or in-vehicle vision modules. For the malicious workload, we introduced the T-Rex cryptominer, configured to mine Ravencoin using the KawPoW algorithm. The miner was launched without administrative privileges to simulate a stealth attack in a real-world edge node where security controls may be limited.

To capture system behavior under both normal and compromised conditions, we used two telemetry tools. The first was nvidia-smi, which logged metrics like GPU utilization, memory usage, and power draw. The second was Nsight Compute, which profiled kernel-level behavior, including streaming multiprocessor throughput, memory bandwidth, and kernel execution time. Data was collected under two conditions: YOLOv8-only (benign), and YOLOv8 with the miner running in the background.

**Results and Observations**
We observed a noticeable drop in system performance when the cryptominer started running in the background alongside the video processing workload. With only YOLOv8 active, the GPU maintained a stable frame rate of about 28 frames per second, and power usage remained around 65 watts. After the T-Rex miner was launched, the frame rate dropped to 14 frames per second, cutting the processing speed in half (**Figure 1**). This reduction in throughput directly affects the ability of a transportation system to process real-time camera input. The frame rate drop is shown clearly in the plot, where performance begins to fall immediately after the miner starts.

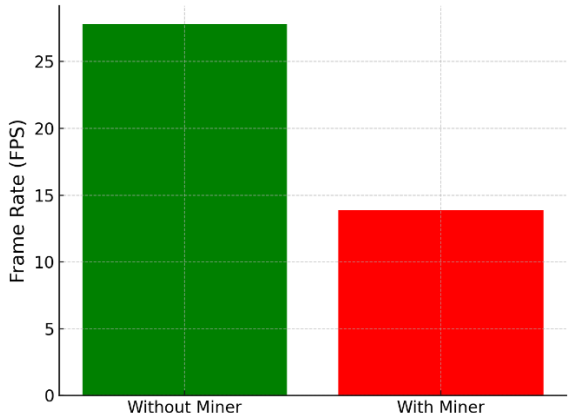

**Figure 1: Average Frame Rate: With vs Without Miner**

At the same time, power consumption increased to around 95 watts. The second plot (**Figure 2**) illustrates this power rise, showing how the miner creates sustained energy demand. GPU utilization also rose sharply and remained at 99 percent, meaning the miner consumed nearly all available processing resources. Memory usage increased from approximately 2800 MiB to 3900 MiB, which could reduce available memory for critical applications.

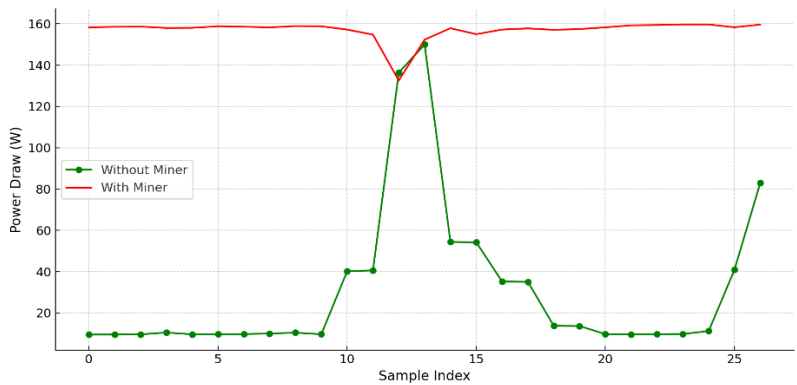

**Figure 2. GPU Power Usage: With vs Without Miner**

Data collected from Nsight Compute supported these observations. **Table 2** below shows a summary of GPU metrics affected by the miner.

**Table 2. GPU Behavior Comparison: Normal Workload vs. Crypto Miner Interference**

| Metric | YOLOv8 Only | YOLOv8 + Miner | Impact |
|---|---|---|---|
| Frame Rate (FPS) | ~28 | ~14 | Real-time processing halved |
| Power Draw (W) | 65 | 95-159 | 30-90% increase |
| GPU Utilization (%) | 40% | 99% | Saturated compute |
| Memory Used (MiB) | ~2800 | ~3900 | Resource hogging |
| sm throughput (%) | Moderate | High + flat | Compute stall |
| dram throughput (%) | Low | Elevated | Memory pressure |

The sm-throughput metric increased significantly and remained constant during mining, showing high kernel activity. Kernel duration also became longer, which indicates that the miner was using the GPU more intensively. DRAM throughput increased as well, meaning the miner placed greater stress on memory bandwidth. A small drop in SM frequency suggested that the GPU was running hotter and possibly throttling to control temperature.

These results raise concerns about GPU safety in transportation. In real-time systems like connected vehicles or roadside units, even a small delay in processing can mean a missing object or a slower response to a hazard. Since the GPU remains active, but overloaded, traditional system checks may not detect the problem. This creates a silent risk that affects both performance and functional safety.

**Detection of the Miner**
To understand whether the presence of a crypto miner can be detected through profiling, we used GPU telemetry logs collected during both benign and miner-active sessions. The goal was to see if a machine learning model could distinguish between normal workloads and hidden mining activity based on the patterns observed in hardware-level metrics. We extracted features such as streaming multiprocessor throughput, DRAM throughput, kernel duration, power usage, and frequency from Nsight Compute reports.

We trained multiple lightweight classifiers on this data. Table 3 shows the detection results for four classifiers - random forest, gradient boosting, linear regression and a neural network with two hidden layers. All the models achieved very high accuracy.

**Table 3: Accuracy of Detection Models**

| Detection Model | Accuracy (%) | Precision (%) | Recall (%) | F1 Score (%) |
|---|---|---|---|---|
| Random Forest | 100 | 100 | 100 | 100 |
| Gradient Boosting | 100 | 100 | 100 | 100 |
| Linear Regression | 100 | 100 | 100 | 100 |
| Neural Network | 100 | 100 | 100 | 100 |

This level of performance shows that the miner leaves behind a very consistent footprint in GPU telemetry. It does not behave like a typical AI workload and can be separated based on its kernel execution characteristics and overall resource consumption.

This detection was done offline, meaning the classification was not happening in real time. However, the results prove that even with limited access, on-device metrics can be used to identify malicious GPU activity with a high degree of confidence. These models require no modification of system drivers, no privileged access, and no extra hardware. In an edge computing context, this makes them suitable for practical deployment in safety-critical transportation systems.

The experiment shows that a stealthy miner, though quiet and non-intrusive on the surface, leaves clear signals when the right telemetry is collected and analyzed. These signals can be used to build a lightweight but effective defense layer in intelligent transportation environments.

**ITS Safety Implications**
The performance degradation caused by the crypto miner has serious implications for intelligent transportation systems. When the GPU is overloaded by unauthorized compute tasks, its ability to support real-time perception and decision-making is reduced. In our case study, the drop-in frame rate and increased latency would delay or entirely miss critical detections in a video pipeline. For example, a vehicle or roadside unit relying on YOLOv8 for pedestrian detection might fail to respond in time if the frame rate is cut in half.

*Puspa et al.*

This becomes a safety issue, not just a performance one. Missed detections or delayed responses in real-world conditions can lead to collisions, near misses, or failure to act in a dynamic traffic situation. In a roadside unit, the system might not detect a vehicle running a red light or a pedestrian crossing unexpectedly. In an autonomous vehicle, a delayed object classification could interfere with braking or navigation decisions. All of this happens silently, without triggering alarms, because the system sees the GPU as functional.

Crypto miners operate in the background and consume resources without causing direct application crashes. This makes them harder to detect and more dangerous in safety-critical environments. Without access to GPU-level observability, transportation systems are left blind to these threats. The signals collected in this case study show that it is possible to identify these risks before they result in visible failures. However, if telemetry is ignored or unavailable, the system continues to run in a degraded and potentially unsafe state.

This highlights the need to treat GPU-level observability as part of the cybersecurity and functional safety framework for connected and automated transportation systems. GPUs are no longer auxiliary accelerators. They are core components of the AI stack, and any compromise to their availability puts the entire perception pipeline at risk.

**CONCLUSIONS**

This paper highlights GPU security as a critical blind spot in intelligent transportation systems. As AI-driven perception becomes central to safety and decision-making, the GPU has evolved from a performance booster to a core operational component. Our case study showed how stealthy crypto mining can silently degrade GPU performance, cutting frame rates in half and increasing power draw, without crashing the system or raising immediate alarms. These impacts pose serious risks to applications like pedestrian detection, traffic monitoring, and autonomous navigation.

We demonstrated that GPU telemetry, collected using existing tools like nvidia-smi and Nsight Compute, is sufficient to detect such misuse with high accuracy. Simple classifiers trained on kernel-level metrics achieved perfect performance, making this approach suitable for deployment on edge nodes. Without requiring deep system access or additional hardware, this method provides a practical path toward improving cybersecurity and functional safety. As transportation systems continue to evolve, GPU observability must become a standard part of system monitoring and protection.

**RECOMMENDATIONS AND FUTURE DIRECTIONS**

This study highlights the importance of monitoring GPU behavior in intelligent transportation systems. As GPUs take on more critical roles in enabling perception and decision-making, overlooking their security creates a significant risk. While most transportation cybersecurity strategies focus on the operating system, network, or CPU, the GPU remains largely unmonitored. This makes it possible for threats like crypto mining to go unnoticed, even while degrading system performance and safety.

We recommend that transportation platforms begin to adopt GPU telemetry monitoring as part of regular system checks. Tools like nvidia-smi and Nsight Compute already provide access to useful performance metrics that can be easily observed and used to monitor different malicious activities. These tools can be used to detect unusual behavior such as sustained power draw, constant utilization, or irregular memory usage. Including GPU performance metrics in the same monitoring pipelines that already track CPU or network activity would offer a more complete view of system health.

The detection models we used in this paper were built using simple classifiers with a small number of features. Despite this, they achieved perfect results in distinguishing normal workloads from mining activity. These models are lightweight and can be integrated into background services that alert the system when GPU usage becomes suspicious. While our tests were performed offline, the telemetry required for detection can be collected in real time. Future implementations could involve live monitoring systems that take action, such as stopping a suspicious process or triggering a fail-safe mode.

There is also a need to extend this work to a wider range of platforms. Many edge devices in transportation use embedded GPUs like the NVIDIA Jetson series, which may have different constraints and fewer accessible metrics. Building similar detection strategies for those platforms is essential if we want broad protection across connected intersections, roadside units, and in-vehicle systems. Linux-based deployments should also be tested, as many field systems do not use Windows.

Although this paper focuses on crypto mining, other GPU-based threats are possible. Attackers could try to steal AI models, introduce adversarial behavior, or use GPUs for hidden data processing. These threats also rely on GPU accessibility and would likely affect similar metrics. Monitoring changes in utilization patterns or kernel behavior could help detect these forms of misuse as well.

We suggest that transportation safety standards begin to include GPU observability as part of overall system readiness. If a vehicle or roadside unit cannot process video in real time, it may fail to detect a pedestrian or recognize a stop sign. These risks cannot be ignored. Setting clear thresholds for frame rate, latency, and resource usage will help identify when a system is operating outside of safe conditions. Without these checks, GPUs can be compromised quietly while the rest of the system continues to appear healthy.

Going forward, stronger collaboration is needed between transportation operators, researchers, and GPU vendors. Sharing the best practices, benchmark datasets, and monitoring tools would help make GPU security more accessible. This paper offers a starting point and a practical example, but the long-term goal should be to treat GPU observability as a standard part of designing and securing transportation systems.

**ACKNOWLEDGMENTS**

This work is based upon the work supported by the National Center for Transportation Cybersecurity and Resiliency (TraCR) (a U.S. Department of Transportation National University Transportation Center) headquartered at Clemson University, Clemson, South Carolina, USA. Any opinions, findings, conclusions, and recommendations expressed in this material are those of the author(s) and do not necessarily reflect the views of TraCR, and the U.S. Government assumes no liability for the contents or use thereof.

**AUTHOR CONTRIBUTIONS**

The authors confirm their contribution to the paper: S. N. Puspa, M. Chowdhury. All authors reviewed the results and approved the final version of the manuscript.